\documentclass[12pt]{iopart}
\usepackage{epsfig}
\usepackage{citesort}
\usepackage{color}
\usepackage{psfrag}
\newcommand{\gtrsim}{\,\rlap{\lower3.7pt\hbox{$\mathchar\sim$}}
\raise1pt\hbox{$>$}\,}
\newcommand{\lesssim}{\,\rlap{\lower3.7pt\hbox{$\mathchar\sim$}}
\raise1pt\hbox{$<$}\,}

\usepackage{color}
\definecolor{Black}{named}{Black}
\definecolor{Blue}{named}{Blue}
\definecolor{Red}{named}{Red}

\newcommand{\OP}{\omega_\mathrm{_P}}

\renewcommand\({\left(}
\renewcommand\){\right)}
\renewcommand\[{\left[}
\renewcommand\]{\right]}

\newcommand{\be}{\begin{equation}}
\newcommand{\ee}{\end{equation}}
\def\bea{\begin{eqnarray}}
\def\eea{\end{eqnarray}}

\begin{document}

\hfill  DESY-09-071 , MPP-2009-58

\title{Constraining  resonant photon-axion conversions in the
Early Universe}
\author{Alessandro Mirizzi}
\address{ Max-Planck-Institut f\"ur Physik (Werner Heisenberg
Institut) \\ F\"ohringer Ring 6,
80805 M\"unchen, Germany}
\author{Javier Redondo}
\address{Deutsches Elektronen Synchrotron \\
Notkestra\ss e 85, 22607 Hamburg, Germany}
\author{G\"unter Sigl}
\address{II. Institut f\"ur theoretische Physik, Universit\"at Hamburg,
Luruper Chaussee 149, 22761 Hamburg, Germany}
\begin{abstract}
\noindent
{The presence of a primordial magnetic field
would have induced resonant conversions
between photons and axion-like particles (ALPs)
during the thermal history of the Universe. These
conversions would have distorted the blackbody spectrum of the 
cosmic microwave background (CMB).
In this context, we derive bounds on the photon-ALP
resonant conversions using the high precision CMB
spectral data collected by the FIRAS instrument on board of the Cosmic Background Explorer.
We obtain upper limits  on the product of the photon-ALP coupling constant $g$ 
times the magnetic field strength $B$ down to $g B \lesssim 10^{-13}$~GeV$^{-1}\,$nG
for ALP masses below the eV scale.}
\end{abstract}
\maketitle
\section{Introduction}
The possible existence of a primordial magnetic field of cosmological origin
has been the subject of an intense investigation during the last few decades.
Despite these efforts, no astrophysical evidence has been reported so far
concerning
magnetic fields over cosmological scales, and only upper limits are reported.
When scaling the original bound from the Faraday effect of distant
radio sources~\cite{Kronberg:1993vk,Grasso:2000wj}
to the now much better known baryon density measured by the Wilkinson
Microwave Anisotropy Probe (WMAP)~\cite{Hinshaw:2008kr}, one has
$B \lesssim2.8\times10^{-7} (l_c/{\rm Mpc})^{-1/2}\,{\rm G}$,
coherent on a scale $l_c\simeq1\,$Mpc~\cite{Blasi:1999hu}.
A recent analysis of the WMAP 5-year data on the Faraday rotation of the linear polarization of the cosmic microwave background (CMB)  gives comparable upper limits ranging from $6 \times 10^{-8}$ to $2 \times 10^{-6}$ G~\cite{Kahniashvili:2008hx}.

The presence of a primordial magnetic field would inevitably produce
resonant  conversions between photons and axion-like particles
(ALPs) in the Early Universe.
This effect was pointed out at first  in a work by Yanagida and
Yoshimura~\cite{Yanagida:1987nf}.
They realized that such a conversion could produce a sizable distortion in the
CMB spectrum.
Since  at that time data indicated a distortion in the Wien region of the CMB, these authors speculated
that a resonant axion-photon conversion could be an intriguing explanation for such an effect.
Nowadays, the blackbody nature of the CMB spectrum has been measured
with a precision better than 1 part in $10^4$ by the
Far Infrared Absolute Spectrophotometer (FIRAS) on board
of the Cosmic Background Explorer (COBE)~\cite{Fixsen:1996nj,Mather:1998gm}.
The astonishing accuracy of such a measurement now allows
to constrain exotic scenarios which would deplete the CMB spectrum, such as hidden photons~\cite{Jaeckel:2008fi,Mirizzi:2009iz},
axions~\cite{Mirizzi:2005ng}, radiative neutrino decays~\cite{Mirizzi:2007jd} or millicharged
particles~\cite{Melchiorri:2007sq}. Recent bounds on photon-ALP mixing using the
precision CMB data~\cite{Mirizzi:2005ng} have focused only on
non-resonant conversions in the late Universe. However, for values of ALP masses allowing resonant conversions during the expansion of the Universe one expects stronger bounds
due to this resonant enhancement.
For this reason, in the present paper we determine the bounds from resonant photon-ALP
conversions in light of the COBE precision CMB data. Moreover, we make use of an accurate
description of the evolution of the cosmological plasma in which the resonant conversions arise,
following the line of our recent study on resonant photon-hidden photon
conversions in the Early Universe~\cite{Mirizzi:2009iz}.

Often, strong bounds on the photon-ALP coupling arise from stellar evolution~\cite{Raffelt:1996wa}.
Our bounds are difficult to compare with these other constraints since they strongly depend on the strength of the primordial fields. If fields near the experimental bounds $\sim 100$ nG  are realized, our bounds can be much stronger than stellar evolution
constraints, but they could also be weaker if the primordial fields were much smaller.
However, in general, our bounds are complementary to stellar evolution ones, since some ALP
models~\cite{Masso:2005ym,Jaeckel:2006xm,Masso:2006gc,Brax:2007ak,Mohapatra:2006pv}
predict a suppression of the photon-ALP coupling in dense stellar interiors, which would not happen in the
relatively diluted primordial plasma. For this reason, we think worthwhile
to explore which independent constraints are  achievable from cosmological
arguments.

The plan of our work is as follows.
In Section~2 we review the mechanism of mixing between photons and
axion-like particles and we present our analytical prescription
to calculate the resonant photon-ALP conversion probability in the expanding
Universe.
In Section~3  we describe our simplified model for the effective photon mass
induced by the primordial plasma.
In Section~4 we discuss the effect of the random primordial
magnetic fields on the resonant conversions and we characterize the averaging
of the conversion probability over the sky and over
the photon polarization,  relevant to study
ALP effects on the CMB monopole spectrum.
In Section~5 we describe the constraints coming from spectral CMB distortions
for ALP masses undergoing resonant conversions after the recombination epoch.
In Section~6 we extend our limits to pre-recombination resonances,
using the experimental limits on the chemical potential $\mu$
of the CMB spectrum
and the agreement between Big Bang Nucleosynthesis (BBN) and the CMB on the effective
number of additional relativistic species at decoupling.
Finally, in Section~7 we comment on the complementarity of our bound
with other astrophysical and experimental constraints and present our conclusions.

\section{Photon mixing with axion-like particles}\subsection{Axion-like Particles}

Axion-like particles (ALPs) are scalar or
pseudoscalar bosons $\phi$ that couple to two photons
with one of the following interaction Lagrangians~\cite{Masso:1995tw}
\bea
{\cal L}_{\rm pseudoscalar} = -\frac{1}{4}g_{-} F_{\mu\nu}\widetilde
F^{\mu\nu} \phi = g_{-} {\bf B\cdot E}\,\phi \,\ ,
\\
{\cal L}_{\rm scalar} = \frac{1}{4}g_{+} F_{\mu\nu} F^{\mu\nu} \phi =
\frac{1}{2}g_{+}  ({\bf B}^2-{\bf E}^2)  \phi\ \,\ ,
\eea
where $\widetilde F^{\mu\nu}= \epsilon^{\mu\nu\alpha\beta} F_{\alpha\beta}/2$ is the dual of the field strength tensor and $\bf E$ and $\bf B$ are the electric and magnetic fields, respectively.
In the presence of a constant external magnetic field one can decompose the electromagnetic field into an
external component and the dynamical part representing photons as
${\bf B}\to {\bf B}^{\rm ext}+\nabla\times \bf A$.
The terms in the Lagrangian containing ${\bf B}^{\rm ext}$
act as mass mixing terms between the axion-like particle $\phi$ and the photon field $A_\mu=(A_0;\bf A)$.
Choosing the radiation gauge $\nabla\cdot {\bf A}=0$ the mixing
terms become evident,
\bea
\label{eq:lagrangian1}
{\cal L}_{\rm pseudoscalar} &=& g_{-} {\bf B}^{\rm ext} \cdot \partial_0 {\bf A}\,\phi +... \ ,
\\
{\cal L}_{\rm scalar} &=&  g_{+} {\bf B}^{\rm ext}\cdot (\nabla\times {\bf A}) \phi  + ... \ .
\label{eq:lagrangian2}
\eea
In our study we will consider the mixing of photons with these particles.
If we write ${\bf A}$ as a plane wave with frequency $\omega$ and wave vector $\bf k$
we can explicitly evaluate its time and spatial derivatives. Reabsorbing a factor of $i$ into $\bf A$
and using $\omega \simeq |\bf k|$ we find that the  mixing part in the
Lagrangian  reads
\bea
{\cal L}^{\rm mix}_{\rm pseudoscalar} &=& g_{-} B_T \omega \, A_{||} \phi \ ,
\\
{\cal L}^{\rm mix}_{\rm scalar} &=&  g_+ B_T \omega \, A_\perp \phi \, ,
\eea
where $B_T$ is the component of the external magnetic field perpendicular to the propagation
direction of photons ($\bf k$)
and  $A_{||},A_\perp$ are respectively the components of $\bf A$ parallel and perpendicular to that
component.
Note that the component of ${ \bf B}^{\rm ext}$ parallel to $\bf k$
does not contribute to the mixing and also that pseudoscalar
fields mix with the parallel component $A_{||}$, while scalars mix with $A_\perp$.
In the following, we denote with $g$ the ALP-photon coupling constant and
consider only the photon component, $\gamma$, that mixes with $\phi$.

Due to the effective mass-mixing in the external magnetic field,
the propagating eigenstates \emph{in vacuum}
are now rotated with respect to $(\gamma,\phi)$ by an angle $\theta$ given
by~\cite{Raffelt:1987im}
\begin{eqnarray}
\label{theta0}
\sin 2\theta &=& \frac{2g B \omega}{\sqrt{m_\phi^4 + (2g B \omega)^2}}
\,\ ,  \\
\cos 2\theta &=& \frac{m_\phi^2}{\sqrt{m_\phi^4+(2g B \omega)^2}}  \,\ ,
\end{eqnarray}
where, for simplicity, we have indicated the transverse component of
the magnetic  field with $B$, and
$m_{\phi}$ is the ALP mass.
This misalignment is well known to produce $\gamma\leftrightarrow \phi$
oscillations with a wavenumber given by~\cite{Raffelt:1987im}
\be
\label{Deltamphi}
k =  \frac{\sqrt{m_\phi^4+(2g B \omega)^2}}{2\omega}  \equiv \frac{\Delta m^2}{2\omega}  \  \  .
\ee
\subsection{Medium effects and resonant MSW conversions}
Photon oscillations into ALPs are modified by the refractive properties of the medium.
In the primordial plasma, photons acquire a non-trivial dispersion
relation which can be parametrized by adding an effective photon mass
$m_{\gamma}$ to the Lagrangian.
This is generally complex, reflecting the absorption properties of the plasma.
However, for most of the parameter space studied here it turns out that absorption is negligible.
Moreover, even when this is not the case our results will show no dependence on it.
Therefore, we have chosen to neglect it in the following exposition and include a few remarks when relevant.
In this case, the \emph{effective} mixing angle $\tilde{\theta}$ is related to the vacuum one by~\cite{Raffelt:1996wa}
\begin{eqnarray}
\label{eq:theta}
\sin 2\tilde\theta &=& \frac{\sin 2\theta}{\[\sin^22\theta+
\(\cos2\theta-\xi\)^2\]^{1/2}} \,\ ,  \\
\cos 2\tilde\theta &=& \frac{\cos2\theta-\xi}
{\[\sin^22\theta+\(\cos2\theta-\xi\)^2\]^{1/2}} \,\ ,
\end{eqnarray}
where the parameter $\xi$ which measures
the significance of the medium effects reads
\be
\label{eq:xi}
\xi  = \frac{m_\gamma^2}{\sqrt{m_\phi^4+(2 g B \omega)^2}}=\cos 2\theta \(\frac{m_\gamma}{m_\phi}\)^2 \ \ .
\ee
As the Universe expands, eventually the condition
\begin{equation}\label{eq:resonance}
m_{\phi} = m_{\gamma} \,\ ,
\end{equation}
is satisfied and $\tilde\theta \to \pi/4$.
When this condition is fulfilled, resonant photon-ALP conversions are possible, analogous to the well-known
Mikheev-Smirnov-Wolfenstein (MSW) effect for neutrino flavor
transitions~\cite{Wolfenstein:1977ue,Mikheyev:1985aa,Mikheyev:1985bb}.
If the photon production and detection points are separated
by many oscillation lengths from a resonance, the oscillation
patterns wash out. Thus, the
transition probability is given by~\cite{Parke:1986jy}
\be
P_{\gamma\to\phi} \simeq \frac{1}{2}+\(p-\frac{1}{2}\)
\cos 2\theta_0\cos 2\tilde\theta\,,
\label{P}
\ee
where we have assumed  the mixing angle $\theta_0$ at the
detection  in vacuum,
$\tilde\theta$ is the effective mixing angle at the
production point (considered to be
at high density) and $p$ is the level crossing probability.
This latter takes into account the deviation from adiabaticity of
photon-ALP conversions in the resonance region.
In particular, one has $p=0$ for a completely adiabatic transition
and $p=1$ for an extremely nonadiabatic one.
The crossing probability $p$ for photon-ALP resonant conversions
can be obtained using the  Landau-Zener expression~\cite{Mirizzi:2009iz}
\begin{equation}
p \simeq \exp\(- {2 \pi r k \sin^2 \theta_r}\) \,\ ,
\label{eq:landau}
\end{equation}
where $k$ is again the $\gamma\to\phi$ \emph{vacuum} oscillation
wavenumber in Eq.~(\ref{Deltamphi}), $\theta_r$ is the vacuum
mixing angle at the resonance and
\begin{equation}
r=\left|\frac{d \ln m^2_\gamma(t)}{dt}\right|^{-1}_{t=t_{\rm res}}
\label{eq:r}
\end{equation}
is a scale parameter to be evaluated at the location where
a resonance occurs.%
\footnote{Our problem is different to the widely discussed of neutrino oscillations or
the photon-hidden photon case studied in~\cite{Mirizzi:2009iz} since here even the vacuum
mixing evolves in time. However, the corrections to Eq.~(\ref{eq:landau}) that we are neglecting
are  unimportant, since the evolution of the vacuum mixing
$\theta$ is very  smooth near the resonance, contrarily to the evolution
of $\tilde \theta$.}

Note that the mixing angle $\theta$ changes as the Universe evolves
because the frequency $\omega$ and the magnetic field $B$
are functions of the redshift.
Denoting with $\omega_0$ and $B_0$ their values at $z=0$,
the frequency grows as $\omega=\omega_0(1+z)$ while the magnetic field
evolution is model-dependent.
In this paper we have decided to focus in the most widely studied case
of considering magnetic fields frozen into the medium, for which $B=B_0(1+z)^2$~\cite{Grasso:2000wj}.
It is clear from Eq.~(\ref{theta0}) that at early times the vacuum mixing angle
becomes maximal, $\sin 2\theta\to 1$.
However, the effective mixing angle $\tilde \theta$ remains small because of the suppression by matter effects.
In fact, at early times, before recombination, $m_\gamma^2$
is given by the plasma frequency $\OP^2$ which is proportional to the free electron density which
scales as $(1+z)^3$. Therefore, at high redshifts
($z\to \infty$), $\xi$ tends indeed to a \emph{constant} given by
\begin{eqnarray}
\label{satXI}
\xi \to & \frac{m_\gamma^2}{2 g B \omega} =
\frac{m_\gamma^2(z=0)}{2 g B_0\omega_0}
\simeq & 2.7 \times 10^{5}
\(g_{10} B_{0,{\rm nG}}\)^{-1}
\left(\frac{T_0}{\omega_0}\right)  \ ,
\end{eqnarray}
where we have used $m_\gamma(z=0)= 1.59\times 10^{-14}$ eV~\cite{Mirizzi:2009iz},
normalized the  photon frequency today $\omega_0$ with respect to the actual CMB
temperature  $T_0=2.725~K$~\cite{Mather:1998gm} and finally defined $g_{10}=g/ 10^{-10}$ GeV$^{-1}$ and $B_{0,{\rm nG}}=B_0/1$
nG as typical values close to their current experimental limits.
For such large value of $\xi$ we can take
$\cos2\tilde\theta\simeq -1$ at the production point in Eq.~(\ref{P}).

Moreover, the FIRAS sensitivity will allow us to bound $P_{\gamma\to\phi}\lesssim 10^{-4}$
which also excludes non-resonant vacuum oscillations with $\sin^22\theta_0\gtrsim10^{-4}$.
Considering resonant conversions we will constrain much smaller mixings so that we can
also take $\cos2\theta_0\simeq 1$ in Eq.~(\ref{P}). This then simplifies to
\begin{equation}
\label{1-p}
P_{\gamma \to \phi} \simeq 1-p \,\ .
\end{equation}
The FIRAS bounds thus require a strongly non-adiabatic resonance.
Under this condition, we can approximate the conversion probability as
\begin{equation}
P_{\gamma \to \phi} \simeq 2 \pi r k \sin^2 \theta_r
\,\ .
\label{proba}
\end{equation}
The product $r k$ is basically the ratio of the characteristic expansion
time of the universe and the
vacuum $\gamma\to\phi$ oscillation length, and for the region
of parameters considered here is large.
Since FIRAS constrains $P_{\gamma\to\phi}\lesssim 10^{-4}$,
this tells us that the vacuum mixing angle $\theta_r$  at the resonance has
to be extremely small.
Therefore, we will work in the \emph{small mixing regime}, for which $\Delta m^2\simeq m_\phi^2$ and
\begin{equation}
\sin2\theta_r\simeq 0.92 \times10^{-5}\,(1+z)^3
\(g_{10} B_{0,{\rm nG}}\)
\,\left(\frac{\omega_0}{T_0}\right)
\,\left(\frac{10^{-14}\,{\rm eV}}{m_\phi}\right)^2\,\ ,
\label{eq:theta0_2}
\end{equation}
where in Eq.~(\ref{theta0}) we have approximated
$\theta_r \simeq gB\omega/m_\phi^2$, since
at the resonance
$m_\phi= m_\gamma$ and  $m_\phi^2\gg 2gB\omega$ for the values of ALP masses we are considering.
Under, this approximation, the expression for
the conversion probability is exceedingly simple, i.e.
\begin{equation}
P_{\gamma \to \phi} \simeq
\frac{ g^2 B^2\pi r \omega }{m_\phi^2} \,\ .
\label{prob}
\end{equation}

%
\section{Cosmological $m_\gamma$ profile}

In order to calculate the conversion probability
we need the profile of the photon effective mass $m_\gamma$ along the cosmological line of sight.
In this work we use the same prescription of our Ref.~\cite{Mirizzi:2009iz} in which
we studied the resonant conversions between photons and
hidden photons in the expanding Universe.

\begin{figure}[t]
\begin{center}
{
\psfragscanon
\psfrag{z}[][l]{$z$}
\psfrag{m}[][l]{$m_{\gamma}$ [eV]}
\includegraphics[width=8cm]{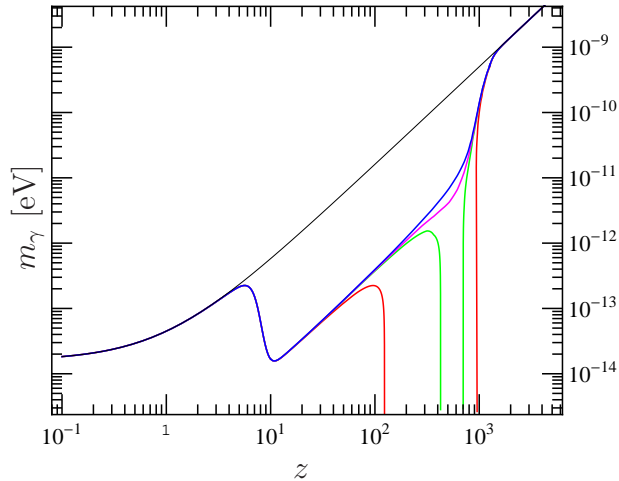}}
\caption{Effective photon mass as function of redshift. The thin line
uses $X_e=1$, the blue, magenta, green and red lines are for $\omega/T=1,3,4,10$,
respectively. These values are in the frequency range probed by FIRAS.
The two sharp dips in the green and red lines bound the region where $m_\gamma^2$ becomes negative.
(See~\cite{Mirizzi:2009iz} for further details)}
\label{Xe}
\end{center}
\end{figure}

The effective mass can be parametrized as
\be
m_\gamma^2 = \OP^2(X_e) \times\[1-0.0073 \(\frac{\omega}{\rm eV}\)^2 \(\frac{1-X_e}{X_e}\)\]\ ,
\ee
where $\OP(X_e) \simeq 1.59\times 10^{-14}(1+z)^{3/2}X^{1/2}_e$~eV is the plasma frequency with $X_e(z)$
the hydrogen ionization fraction as a function of redshift,
which we take from~\cite{Seager:1999bc}.
We note that the CMB energies probed by FIRAS are much smaller than
the first excitation energies of the hydrogen
(when the temperatures are low enough to allow bound atoms)
so that we have neglected these effects in the expression of the effective mass.
We have also ignored  the subleading helium contribution.
Based on~\cite{Mirizzi:2009iz},  in Fig.~\ref{Xe} we show our reference profile for the cosmological history
of  the effective photon mass $m_\gamma$ as a function of redshift $z$.
We recall that the evolution of the effective photon mass which depends on the history of the
ionization fraction $X_e(z)$ and also on the photon energy, is extremely complex.
Above a temperature $T\sim 0.5$ eV (redshift $z\sim 1100$) hydrogen
is fully ionized. As the Universe temperature decreases,
photons cannot ionize hydrogen efficiently and electrons and protons slowly combine.
This makes the Universe very transparent to radiation, indeed
releasing the photon bath which we see today as the CMB.
This is the so-called recombination epoch. Later on ($z\lesssim 6$) the Universe becomes ionized again due to ultraviolet radiation from the first
quasars or population III stars.
These effects produce the non-monotonic behavior in the shape of the
effective photon mass as a function of
redshift seen in Fig.~\ref{Xe}. Note also that during recombination the effective mass squared can
become negative if the photon energy is large enough.

Along such a non-monotonic mass profile multiple resonances can occur for a given ALP mass 
(provided $m_{\phi} \lesssim 10^{-12}$~eV).
If this is the case, Eq.~(\ref{prob}) tells us that the most adiabatic crossing,
quantitatively the more important, is the one that happens at earlier times.
This is particularly clear when we express all red-shift dependent quantities as their
values at $z=0$ times their redshift dependence and use $r\propto (3 H)^{-1}$,
 with $H$ the Hubble constant, 
to find $P_{\gamma\to \phi}\propto (1+z)^5/H$ which increases with redshift since $H$ grows
at most as $(1+z)^2$ during the radiation dominated epoch.

We note that for  photon energies
$\omega/T\gtrsim 3.8$ there is a small period during the dark ages
for which the effective mass squared becomes negative.
This implies that there is one redshift at which $m_\gamma^2=0$
and at which, in principle, ALPs of arbitrarily small mass can be
resonantly produced.
However, due to the fast drop of $m_\gamma^2$, this crossing
is extremely non-adiabatic, as we have explicitly checked.
For this reason, we will not consider this case hereafter.

In summary, in the case of multiple crossings the  most relevant one
happens before reionization.
As a consequence, for $m_{\phi}\gtrsim10^{-14}$ eV  we need
to evaluate resonances only for $z\gtrsim10$.
This situation  simplifies our calculations because after
reionization the plasma is  complicated by the
presence of density inhomogeneities~\cite{Mirizzi:2009iz}.

%
%

\section{Magnetic fields at the resonance and probability averages}

In principle, the primordial magnetic field is not known, although it is generally
expected to show a somewhat turbulent structure~\cite{Mack:2001gc}.
We would like to clarify how it affects the resonant photon-ALP conversions.
For this reason, we have to compare the typical width of a resonance region
with the coherence scale of the magnetic field.
The half-width of the resonance is, according to Eq.~(\ref{eq:theta}),
$\delta \xi(t)= \sin 2\theta_r$, which corresponds to a length scale
\begin{eqnarray}
\tau_r &=&  r \sin2\theta_r\\
&\lesssim& 1.3 \times 10^{-2}\,(1+z)^{3/2}\,
\left(g_{10} B_{0,{\rm nG}} \right)
\,\(\frac{\omega_0}{T_0}\)
\,\(\frac{10^{-14}\,{\rm eV}}{m_\phi}\)^2\,{\rm Mpc}
\nonumber\\
&\lesssim& 0.42
\(g_{10} B_{0,\rm nG} \)
\,\left(\frac{\omega_0}{T_0}\right)
\,\left(\frac{10^{-14}\,{\rm eV}}{m_\phi}\right)\,{\rm Mpc}\ ,
\label{taur}
\end{eqnarray}
where for the first numerical estimate we have used
Eq.~(\ref{eq:theta0_2}) and
$r\lesssim (3H)^{-1} \simeq1.4 (1+z)^{-3/2}\,$Gpc, with $H$ the Hubble constant.
Eq.~(\ref{taur}) results from the fact that for the most relevant resonance before reionization (at $z\gtrsim10$)
one has $m_\phi=m_\gamma\gtrsim10^{-14}\,[(1+z)/10]^{3/2}\,$eV.
The resonance half-width is thus smaller than the coherence length
$l_c\sim l_{c,0}\,(1+z)^{-1}$ of the magnetic field for
\begin{equation}\label{eq:testy}
g_{10}B_{0,{\rm nG}}\lesssim 1.3\times10^{-2}\,\left(\frac{m_\phi}{10^{-14}\,{\rm eV}}\right)^{1/3}
\,\left(\frac{T_0}{\omega_0}\right)
\left(\frac{l_{c,0}}{{\rm Mpc}}\right) \,\ .
\end{equation}

The boundary of our exclusion bounds will satisfy this constraint (see Fig.~\ref{alpFIG})
so that we can consider the magnetic field constant during the resonance to compute it.
At couplings larger  than the ones at
the exclusion boundary, the resonance width will eventually become larger than the
magnetic field coherence length and photons will see different magnetic field domains during
the resonance.
Nevertheless,  the transition probability
increases by increasing the value of $g$, so that we can also exclude this upper region.
Therefore, also in such a situation, we will continue to
use our simple prescription of taking the magnetic field 
evaluated exactly at the crossing point, defined by our Eq.~(\ref{eq:resonance}), 
and constant during the resonance%
\footnote{Possible corrections due  to the variation of $B$ inside of the resonant region
(see, e.g., the discussion in~\cite{Balantekin:1988aq} in the neutrino case)
are not expected to modify our exclusion plot above the  boundary.
}.

\begin{figure}[t]
\begin{center}
{
\psfragscanon
\psfrag{b}[][l][0.9]{$\bf B$}
\psfrag{bt}[][l][0.9]{${\bf B}_T$}
\psfrag{x}[][l]{$\bf x$}
\psfrag{psi}[][l][0.8]{$\psi$}
\psfrag{a1}[][l][0.7]{$A_{||}$}
\psfrag{a2}[][l][0.7]{$A_\perp$}
\psfrag{gam}[][l][0.8]{$\varphi$}
\psfrag{ep}[][l]{$\hat \epsilon$}
\psfrag{post}[][l]{post-recombination}
\psfrag{pre}[][l]{pre-recombination}
\psfrag{las}[][l][0.7]{last scattering surface}
\psfrag{pt}[][l][0.5]{photon trajectories}
\psfrag{r}[][l][0.5]{resonant shell}
\includegraphics[width=10cm]{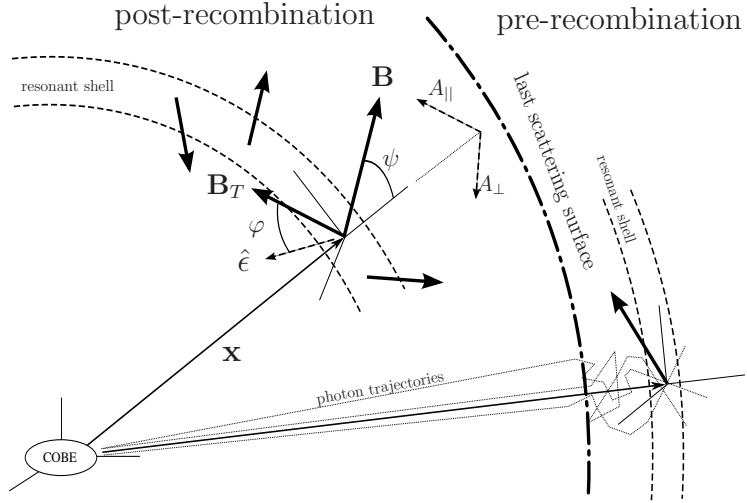}
}
\caption{\small Picture of resonant photon-ALP oscillations of the cosmic microwave radiation.
The transitions happen during a very short time in the history of the universe (the resonant shell) determined by the ALP mass which here is shown confined by dashed lines.
The primordial magnetic field  can be assumed to be constant during the
resonance (the shell thickness) but both its strength and orientation depend on the angular position in the sky.
Resonant transitions happening after recombination deplete CMB photons depending on their polarization.
If  resonances happen before recombination, photon trajectories can form any angle with ${\bf B}$ during the resonance
and then be re-scattered towards the observer such that information about the polarization is lost.
In this case, the CMB light from the last scattering surface is effectively
averaged over the angle $\psi$.}
\label{ske}
\end{center}
\end{figure}

However, the magnetic field direction and strength during the resonance will generally be
different along different directions in the sky. Moreover, the photon-ALP resonant conversions
depend on the relative orientation of the photon polarization
and the magnetic field direction, so that in different magnetic domains
different photon polarization states play the role of
$A_{||}$ and  $A_\perp$, see Fig.~\ref{ske}.
For a generic photon polarization, the $B$ strength entering the conversion probability in Eq.~(\ref{prob}) is%
\footnote{Eq.~(\ref{eq:Bpol}) holds for parity-odd ALPs; when considering
parity-even one finds
$|{\bf B}({\bf x})\times \hat \epsilon| = |{\bf B}({\bf x}) \sin\psi({\bf x}) \sin \varphi |$ instead.}
\be\label{eq:Bpol}
B= |{\bf B}({\bf x})\cdot \hat \epsilon| = |{\bf B}({\bf x}) \sin\psi({\bf x}) \cos \varphi |\ ,
\ee
where ${\bf x}$ is the position vector of the resonance region in a particular direction $\hat x$,
$\hat \epsilon$ is the photon polarization vector ($|\hat \epsilon|=1$,$\hat \epsilon\times \hat x=0$),
$\psi({\bf x})$ is the angle between the magnetic field ${\bf B}({\bf x})$ and the photon propagation direction $\hat x$ and
$\varphi$ the angle between
${\bf B}_T$ (the component of the magnetic field perpendicular to $\hat x$) and $\hat \epsilon$, see Fig.~\ref{ske}.

In this paper we will mainly discuss bounds of photon-ALP mixing from the CMB monopole.
If the resonance happens after recombination,
the full-sky and polarization averaged photon-ALP conversion
probability Eq.~(\ref{prob}) is
\be
\langle P_{\gamma\to \phi } \rangle =\frac{\pi g^2 r \omega}{m_\phi^2}
\frac{1}{3}\langle B^2 \rangle \,\ ,
\label{eq:averprob}
\ee
where we have defined a conveniently normalized average
\begin{eqnarray}
\langle B ^2\rangle &\equiv&
3 \int \frac{d \Omega_{\hat x}}{4 \pi}\int_0^{2 \pi}
\frac{d\varphi}{2\pi}  |{\bf B}|^2 \sin^2\psi \cos^2\varphi
\nonumber \\
&=&
\frac{3}{2} \int \frac{d \Omega_{\hat x}}{4 \pi} |{\bf B}|^2 \sin^2 \psi \ ,
\label{eq:averMAG}
\end{eqnarray}
when we recall that implicitly ${\bf B}$ and $\psi$ depend on ${\bf x}$.
The convenience of this average is clarified in two simple cases:
a) the absolute value $|{\bf B}|$ is constant in the whole sky but has
a random direction and b) it has a fixed direction. In both cases we find
\be
\langle  B^2\rangle =  |{\bf B}|^2   \,\ .
\label{eq:averMAG_simp}
\ee

When the resonance happens before recombination,
the photon trajectories
can form any angle $\psi$ with ${\bf B}$ during the resonance and
then be re-scattered towards the observer such that information about the polarization is lost.
Since the photon trajectories are not straight,
the $\psi$ angle is not correlated with
the magnetic field direction,
photon polarization and the direction in the sky.
Because of this,
one has to perform a local  average over
$\psi$ in each magnetic domain before doing the sky-average, i.e.
\bea
|{\bf B}|^2
\int \frac{d \varphi}{2 \pi}\frac{d \Omega_\psi}{4 \pi}
\sin^2\psi \cos^2\varphi =
\frac{1}{3} |{\bf B}|^2 \ .
\eea
Assuming that ${\bf B}={\bf B}({\bf x})$ does not vary very
much on the distance scales of the horizon at last-scattering surface,
we shall then define a sky average
\be
\langle B^2\rangle = \int \frac{d \Omega_{\hat x}}{4 \pi} |{\bf B}|^2 \,\ ,
\label{eq:averMAG_pre}
\ee
so that also in the pre-recombination epoch one can use Eq.~(\ref{eq:averprob})
for the conversion probability.
Note that if $|{\bf B}|$ is isotropic, also in this case we obtain
$\langle  B^2\rangle =  |{\bf B}|^2$.
This expression neglects in principle decoherence effects from
photon scattering absorption during the resonance.
However as we will see in Sec.~6 these
effects are not relevant if the conversion probability is small.
%
%

\section{Bounds in the post-recombination era}
The CMB spectrum measured by FIRAS fits extremely well to a black-body
spectrum at a temperature $T_0=2.725\pm 0.002$K~\cite{Mather:1998gm}.
The energy range of the  CMB spectrum measured by FIRAS~\cite{Fixsen:1996nj}
is $2.84\times10^{-4}\,{\rm eV}\leq \omega_0 \leq 2.65\times 10^{-3}\,$eV,
corresponding to $1.2 \leq \omega_0/T_0 \leq 11.3$.
In that region, the CMB blackbody becomes unprotected to distortions
below a cosmic temperature $\sim$ 50 eV, which corresponds to a
photon plasma mass of $\sim 10^{-6}$ eV. On the other hand, today
the average plasma mass  for photons $m_\gamma$ is as low as $2\times 10^{-14}$ eV.
If ALPs exist with a mass between these two values they
will be produced resonantly and leave their imprint on the CMB as a
frequency-dependent distortion [see Eq.~(\ref{eq:averprob})].
Bounds for higher masses are considered in the following section.

In order to obtain our  bound, we have considered the distortion of the overall blackbody spectrum.
To this end we use the COBE-FIRAS data for the experimentally measured
spectrum, corrected for foregrounds~\cite{Fixsen:1996nj}.  Note that
the new calibration of FIRAS~\cite{Mather:1998gm} is within the old
errors and would not change any of our conclusions.  The $N = 43$ data
points $\Phi^{\rm exp}_i$ at different frequencies $\omega_{i}$ are
obtained by summing the best-fit blackbody spectrum  to the
residuals reported in Ref.~\cite{Fixsen:1996nj}.  The errors
$\sigma^{\rm exp}_i$ are also available.
In the presence of
photon-ALP conversion, the original intensity of the ``theoretical
blackbody'' monopole at temperature $T$,
\begin{equation}
\label{planck}
\Phi^0({\omega},T) = \frac{\omega^3}{ \pi^2}
\big[ \exp (\omega/T )-1 \big]^{-1} \ ,
\end{equation}
would be deformed to
\be
\Phi({\omega},T,\lambda)=\Phi^0({\omega},T)[1-\langle P_{\gamma\to\phi}\rangle] \ ,
\ee
where  $\langle P_{\gamma\to\phi}\rangle$ is the sky
average of the polarization averaged photon-ALP conversion probability,
defined in Eq.~(\ref{eq:averprob}).
We can then build the reduced chi-squared function
\begin{equation}
\chi_\nu^2(T,\lambda)=\frac{1}{{N}-1}
\sum_{i}^{N}
\bigg[\frac{\Phi^{\rm exp}_i-\Phi({\omega}_i,T,\lambda)}
{\sigma^{\rm exp}_i} \bigg]^2\,.
\end{equation}
We minimize this function with respect to $T$ for each point in the
parameter space  $\lambda=(m_\phi,g \langle B^2\rangle^{1/2})$, i.e.\ $T$ is an
empirical parameter determined by the $\chi_\nu^2$ minimization for each
$\lambda$ rather than being fixed at the standard value.

In Fig.~\ref{alpFIG} we show our exclusion contour.
In particular, the region above the continuous curve is the
excluded region at 95\% C.L., i.e.\ in this region the chance
probability to experimentally obtain larger values of $\chi_\nu^2$ is
lower than~5\%.

We shall stress again that the CMB photons from different angles have traversed magnetic fields during the resonance that have in principle different strengths and directions.
This would induce anisotropies and polarisation.
The difference between a direction parallel and orthogonal to the magnetic field must not exceed the observed
CMB temperature anisotropy, $\Delta T/T<10^{-5}$.
However, the bound achievable from this anisotropy pattern is expected
to be less stringent than the one put from the distortion of the overall
blackbody spectrum~\cite{Mirizzi:2005ng}.
Possible improvement of this bound would require a detailed investigation in terms of  the multipole expansion of the
CMB temperature fluctuations. We deserve this task for a future work.

\section{Bounds  in the pre-recombination era}

The resonant conversion of photons into ALPs produces an energy-dependent depletion of
the CMB which can
be constrained by FIRAS data.
However, if this resonance happens before recombination, the CMB is still coupled
to the primordial plasma and
these distortions can be processed by photon-plasma interactions which will tend to thermalize the spectrum.
Photon scattering and absorption during the resonance
can in principle affect our result Eq.~(\ref{prob})
producing a damping of the photon-ALP conversions.
Both effects can be introduced in an imaginary contribution to the effective photon mass squared.
In the limit of strong damping, the flavor relaxation rate becomes independent on the details of the scattering processes.
When this situation dominates a resonance, taking the results of~\cite{Jaeckel:2008fi} 
(computed in the photon-hidden photon case but valid for a general mixing case)
and translating them into our problem in the relevant small mixing case, we obtain

\be
\label{dampdom}
P_{\gamma\to \phi}\simeq \frac{1}{2} \(1-p^2\) =
\frac{1}{2}\[1-\exp\(-2\frac{\pi g^2 B^2 \omega}{3 H m_\phi^2}\)\] \ ,
\ee
where $p$ is the same crossing probability given by Eq.~(\ref{eq:landau}).
The effects of decoherence become manifest in the
fact that for adiabatic resonances, $p\sim 0$, the system reaches thermal
equilibrium, and photons and ALPs equilibrate their
populations which, of course, results in a transition probability of $1/2$.
However, in the non-adiabatic limit, this expression gives exactly our
Eq.~(\ref{prob}), with $r=(3 H)^{-1}$ in this regime,
so our results do not depend on whether the resonance is vacuum
or damping dominated and we can use Eq.~(\ref{prob}) in both the cases.
For the sake of the simplicity, we neglect possible cases in which the scales 
of the damping and of the resonance are of the same order.
As a summary, we only have to consider the processing of the CMB spectrum \emph{after}
the resonance to compare it with the FIRAS data.

Note that the processing strongly depends on the temperature at which the resonance takes place,
since it determines if the different photon interactions are effective or not.
The response of the primordial plasma to distortions of the Planck distribution is
driven by two fundamental classes of processes: Compton scattering conserves
the number of photons, whereas double Compton scattering, Bremsstrahlung and their time
reversal analogues change the number of photons.

Compton scattering  is the fastest process and is responsible for
the change of photon direction and polarization described already in Sec.~4.
Compton scattering can also be effective at redistributing the energies of photons, however \emph{per se}
cannot change the photon number.
It can provide \emph{kinetic} equilibrium of the photon distribution provided it
is efficient, i.e.,
\be
\int_0^z\frac{n_e \sigma_{\rm Th}}{2 H}\frac{T}{m_e}\frac{dz}{1+z}\gg 1
\quad \to \quad
z\gg 2 \times 10^{5}\ ,
\ee
where  $\sigma_{\rm Th}$ is the Thompson's scattering cross section.
This condition corresponds to masses $m_\phi\gg 1.4 \times 10^{-6}$ eV.
Therefore, below this mass the plasma is so weakly
coupled that it cannot process the distortion.
In such a condition, the treatment given in the previous section holds
with the only difference being the averaging of
photon directions during the resonance, as discussed in Sec.~4.
The corresponding bound is shown in Fig.~3.

For masses $\sim 10^{-6}$ eV the situation is more complex since the
resonance happens when Compton scattering is neither efficient nor inefficient.
The evolution of the distortions has to be studied by numerically evolving the spectrum with
the Kompaneets equation. We believe this is beyond the scope of this paper.

Whenever Compton scattering is efficient, the photon distribution acquires, in short time scale after the
resonance, a Bose-Einstein shape characterized by a chemical potential.
For small distortions, this is given by~\cite{Hu:1992dc}
\be
\mu_{r} = \frac{-1}{2.142}\(3 \frac{\delta \rho}{\rho}-4
\frac{\delta n}{n}\) \simeq -0.05\ g^2_{10} \langle B_{0,{\rm nG}}^2\rangle \,\ ,
\label{mu0}
\ee
where $\delta\rho/\rho$ and $\delta n/n$ are the fractions of the CMB energy and photon number converted into ALPs during the resonance. Note that, in contrast to the case often studied in the literature, resonant photon-ALP conversion absorbs relatively more energy than photon number and therefore produces negative chemical potentials.

\begin{figure}[t]
\centering
{
\psfragscanon
\psfrag{a1}[][l][0.8]{post-recombination}
\psfrag{a2}[][l][0.8]{weak-coupling}
\psfrag{a3}[][l][0.8]{ \hspace{0.2cm}$\mu$}
\psfrag{a4}[][l][0.8]{ \hspace{0.3cm}$\Delta N_{\rm eff}$}
\psfrag{malp}[][l]{$m_{\phi}$ [eV] }
\psfrag{galp}[][l]{$g \langle B^2 \rangle^{^{1/2}}\times 10^{10}$ GeV $\times$ nG\vspace{1cm}}
\includegraphics[width=14cm]{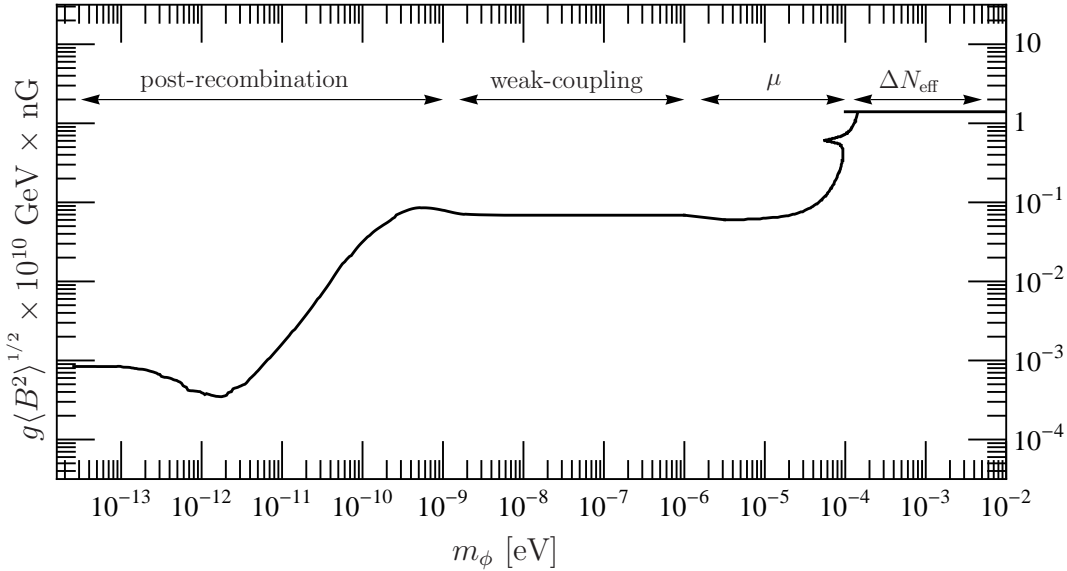}
}
\caption{
Bounds on the ALP parameter space from distortions of the CMB blackbody spectrum
caused by resonant $\gamma\to\phi$ oscillations.
The ALP mass determines the time of the resonant transition.
In the region labeled as ``post-recombination''
the transition happens in the post-recombination epoch
whereas outside this region the transition occurs before recombination and CMB
last scattering. However, in the region labeled as ``weak-coupling''
the transitions happen when the plasma is
so weakly coupled that it cannot process the distortions.
In the region labeled $\mu$, Compton scattering would restore a Bose-Einstein
spectrum with a
chemical potential that can be erased through inverse double Compton scattering
and Bremsstrahlung.
For $m_\phi \gtrsim 0.1$ meV all distortions are erased but the produced ALPs still
contribute to the  cosmic radiation density and can be constrained by comparing the
number of effective number of relativistic species at BBN and CMB decoupling.
Here $g$ is the ALP-photon coupling constant and $ \langle B^2 \rangle^{1/2}$ an
sky average of the comoving magnetic field during the resonance. }
\label{alpFIG}
\end{figure}

Inverse Double Compton scattering (DC) and Bremsstrahlung (BS)
occur much less frequent than Compton scattering
but in contrast to the latter can change the photon number.
Therefore, on a longer time scale they can absorb the necessary photons to recreate a
pure Planck distribution, i.e. they can erase the chemical potential. For small distortions,
the evolution of $\mu$ can be approximated by~\cite{Hu:1992dc}
\bea
\frac{d \mu}{d t} = -\mu\(\frac{1}{t_{\rm DC}}+\frac{1}{t_{\rm BS}}\)\ , \quad \rm  with \\
t_{\rm DC}= 1.06\times 10^{8} {z_6}^{9/2}\ {\rm s}.
\quad ;\quad
t_{\rm BS}= 3.73\times 10^{8} {z_6}^{13/4}\ \rm s\ ,
\eea
where we have taken the values $Y_p=0.25,\Omega_bh^2=0.0223$ for the $\Lambda$CDM model and defined $z_6=z/10^6$.
During radiation domination we can write the expansion time as $t=(2 H)^{-1}\simeq 2.38\times 10^7z_6^{-2}$ s.
We have computed for each ALP mass the time at which the resonance takes place and the evolution of the chemical potential.
The FIRAS data sets the bound on the value of the chemical
potential today $\mu_0<9\times 10^{-5}$, which gives the exclusion range denoted
$\mu$ in Fig.~\ref{alpFIG}.

The bounds from the distortions of the CMB spectrum vanish
very fast for masses above $\sim 0.1$ meV because
Double Compton scattering  and Bremsstrahlung  become very efficient.
However, there is  a further bound we can consider
for resonances happening in the post BBN epoch.
During ALP production a fraction of the energy stored in the CMB is transferred to ALPs, which immediately decouple from the thermal bath.
If the ALPs produced in this way have sufficiently small masses to
be relativistic during the epoch of the CMB formation
(i.e. roughly $m_\phi\lesssim 1$ eV) then they  behave as a non-standard
contribution to the radiation energy density.
Therefore, the radiation energy density measured from CMB anisotropies
would keep track of the ALPs contribution.
A comparison of the radiation energy density during Big Bang Nucleosynthesis (not affected by ALPs resonant production) and CMB decoupling gives
an upper limit on this contribution, namely $x<\delta \rho/\rho=0.2$~\cite{Jaeckel:2008fi,Iocco:2008va}.
In our case this translates into a bound
\be
g_{10} \langle B_{0, {\rm nG}}^2\rangle^{1/2} < 1.4 \ .
\ee
We show our combined constraints in Fig.~\ref{alpFIG}. In this figure we have used the full expression in Eq.~(\ref{dampdom}) since
above $m_\phi \simeq 0.1$ meV the photon-ALP conversion probability bound cannot be considered small (also, for any mass a fraction of $0.2$ of the photons converted into ALPs requires $P_{\gamma\to \phi}$ of order 1).

\section{Discussion and conclusions}

In this paper, we have calculated bounds on photon-ALP oscillations
in the primordial magnetic field, deriving updated constraints from
the high precision CMB spectrum data collected by the FIRAS
instrument on board of COBE.
A previous study~\cite{Yanagida:1987nf}
was derived in the pre-COBE  era and
it lacked a detailed treatment of the effects of the plasma medium
on the photon-ALPs oscillations.
This has motivated us to re-evaluate the bounds.
We obtain limits on the product of the ALP-photon coupling $g$ times the sky and polarization averaged magnetic field
$\langle B^2\rangle^{1/2}$,
\be
g \langle B^2\rangle^{1/2}\lesssim\  10^{-13}\sim  10^{-11}\,\ \textrm{GeV}^{-1}\,\textrm{nG} \ ,
\ee
for ALP masses between $10^{-14}\,$eV and $10^{-4}\,$eV.
Slightly weaker bounds were also derived for higher  ALP masses.

Our bound nicely connects with the one obtained in~\cite{Mirizzi:2005ng}
for ALP masses less
than $10^{-14}$~eV, considering only non-resonant conversions today.
Our argument allows to extend the sensitivity of CMB measurements also in a region
of the parameter space that was previously unconstrained.
We mention that ALPs can also be thermally produced
in Early Universe  via processes like
$e\gamma\to e \phi$, forming a relic background.
The $\phi \gamma\gamma$ vertex would also allow for radiative
decays of these particles. In this case, additional cosmological
bounds can be obtained
for $m_\phi \gtrsim$~eV, as studied in~\cite{Masso:1997ru}.

There are several interesting proposals to measure the CMB spectrum with
higher sensitivity than the FIRAS instrument. The DIMES (Diffuse Microwave Emission Survey)
 proposal~\cite{Kogut:1996zb}  aims
to probe the region of smaller frequencies (2-100 GHz corresponding
 to $0.035\leq \omega_0/T_0\leq 1.8$) with sensitivity
comparable to FIRAS. Unfortunately, as discussed above, the distortions grow
with increasing frequency, so DIMES would not
significantly strengthen bounds on axion-photon oscillations, in contrast to
constraints on hidden photon - photon mixing for
which conversion probabilities are inversely proportional to the photon
energy. For this reason, the proposed FIRAS II~\cite{Fixsen:2002} would
be more relevant for photon-axion mixing constraints since it would shrink
the FIRAS error bars by almost two order of magnitude and
reach much higher frequencies (60-3600 GHz, or $1\leq \omega_0/T_0\leq 65$).
 If such a
project is realised the sensitivity to photon-ALP oscillations will very
likely improve by more than an order of magnitude in
$g \langle B^2\rangle ^{1/2}$.
A slight improvement on our bound in the pre-recombination era
could be achieved thanks to new experiments, like ARCADE (Absolute
Radiometer for Cosmology, Astrophysics, and Diffuse Emission) which would
constrain the chemical potential of the CMB spectrum
$|\mu|$ down to $2 \times 10^{-5}$~\cite{Singal:2008zz}.

Let us recall that recent results from the CAST experiment~\cite{Andriamonje:2007ew,Zioutas:2004hi,Arik:2008mq}
give a direct experimental bound on the ALP-photon coupling of $g \lesssim 8.8 \times
10^{-11}$~GeV$^{-1}$ for $m_{\phi} \lesssim 0.02$~eV,
slightly stronger than the long-standing globular-cluster
limit~\cite{Raffelt:2006cw}. For ultra-light ALPs
($m_{\phi} \lesssim 10^{-10}$~eV) a stringent limit
from the absence of $\gamma$-rays from SN~1987A gives $g \lesssim
1\times 10^{-11}$~GeV$^{-1}$~\cite{Brockway:1996yr} or even $g
\lesssim 3\times 10^{-12}$~GeV$^{-1}$~\cite{Grifols:1996id}.
We stress that without direct evidence for
a primordial magnetic field, our  bounds on $g\langle B^2\rangle^{1/2}$
do not allow to constrain directly the coupling constant $g$.
However, if a primordial magnetic field would be found with
values close to the current upper bound,
the resulting CMB limit on $g$  for $m_{\phi} \lesssim
10^{-4}$~eV would overcome the barrier placed by
current experimental and astrophysical bounds.
Conversely, if  ALPs will be eventually discovered improving the current sensitivity
of the solar axion helioscope CAST, or with new techniques~\cite{Hoogeveen:1990vq,Sikivie:2007qm,Cantatore:2008ju} in laser
experiments like photon regeneration~\cite{ALPS,Fouche:2008jk,Chou:2007zzc,Afanasev:2008jt} or laser polarization~\cite{Cantatore:2008ju,Battesti:2007zz},
our cosmological argument will provide a complementary constrain on
the strength of the primordial magnetic field.

Moreover, we note that our limit applies also to the so called chameleon-like
scalar particles mixing with photons~\cite{Khoury:2003rn,Mota:2006ed}.
Such particles are introduced to explain the acceleration of the Universe, as inflation or dark energy fields~\cite{Brax:2004qh}, or to
cause variations in the fundamental constants~\cite{Mota:2003tm,Olive:2007aj}.
In general, chameleons have properties that depend on the environment. For this reason,
they can evade astrophysical and CAST bounds, since in the dense
stellar environment they  become so heavy that they can not be produced
in the usual reactions~\cite{Brax:2007ak}. In these cases one has thus to rely
on the limits from the laboratory experiments, which take place essentially in vacuum.
For a wide class of chameleon models, the laser PVLAS experiment would rule out values of $g \gtrsim 5 \times
10^{-7}$~GeV$^{-1}$~\cite{Zavattini:2007ee} (see also~\cite{Ahlers:2007st,Brax:2007hi,Gies:2007su,Chou:2008gr}).
Strong bounds also come from observations of starlight polarization~\cite{Burrage:2008ii}.
In the expanding Universe, given the low plasma density, these particles would behave essentially as standard ALPs, so that our bounds can be directly applied to this case.
In this sense, if  primordial magnetic fields will be discovered, our argument
would rule-out the recently proposed mechanism of chameleon-photon conversions
to explain the observed supernovae Ia brightening~\cite{Burrage:2007ew}.

Furthermore, let us point that the kind of analysis performed in this work can be also
extended to other particles having two-photon vertices and sub-eV masses.
For instance, a very similar analysis could be performed to massive spin-2 particles~\cite{Biggio:2006im}
like Kaluza-Klein gravitons~\cite{Deffayet:2000pr}.

As a final remark,  our study is relevant for recent studies of conversions of high-energy gamma rays into
ALPs in the intergalactic magnetic field in relation to the observed cosmic transparency of high energy
gamma sources~\cite{De Angelis:2007yu,De Angelis:2007dy,DeAngelis:2008sk}
or to ultra high-energy cosmic ray propagation~\cite{Csaki:2003ef}.
If eventually a primordial magnetic
field will be discovered, a definitive verdict on the impact of these
fascinating mechanisms would require a combined analysis including our new CMB
constraints.
In this sense, it is nice to realize that our cosmological limits obtained
with microwave photons could have  relevant consequences for
signatures of ALPs in high-energy gamma sources. This confirms once more the
broad range of energies potentially accessible through axion-like particle
searches.

\section*{Acknowledgements}
This work was supported by the Deutsche Forschungsgemeinschaft (SFB 676 ``Particles, Strings
and the Early Universe: The Structure of Matter and Space-Time) and by the European Union
ILIAS (contract No. RII3-CT-2004-506222).
The work of A.M. is supported by the Italian Istituto Nazionale di Fisica
Nucleare (INFN).
We thank Georg Raffelt and Andreas Ringwald for reading the manuscript and for valuable
comments on it. A.M. acknowledges kind hospitality at the Hamburg
University during the development of this work.

\section*{References} 



\begin{thebibliography}{00}

\bibitem{Kronberg:1993vk}
P.~P.~Kronberg,
``Extragalactic magnetic fields,''
Rept.\ Prog.\ Phys.\  {\bf 57}, 325 (1994).
\bibitem{Grasso:2000wj}
D.~Grasso and H.~R.~Rubinstein,
``Magnetic fields in the early universe,''
Phys.\ Rept.\  {\bf 348}, 163 (2001)
[astro-ph/0009061].

\bibitem{Hinshaw:2008kr}
{\bf WMAP} Collaboration, G.~Hinshaw {\it et al.},
``Five-Year Wilkinson Microwave Anisotropy Probe
Observations:Data Processing, Sky Maps, \& Basic Results,''
Astrophys.\ J.\ Suppl.\  {\bf 180}, 225 (2009)
[arXiv:0803.0732 [astro-ph]].

\bibitem{Blasi:1999hu}
P.~Blasi, S.~Burles and A.~V.~Olinto,
``Cosmological Magnetic Fields Limits in an Inhomogeneous Universe,''
Astrophys.\ J.\  {\bf 514}, L79 (1999)
[astro-ph/9812487].

\bibitem{Kahniashvili:2008hx}
T.~Kahniashvili, Y.~Maravin and A.~Kosowsky,
``Primordial Magnetic Field Limits from WMAP Five-Year Data,''
arXiv:0806.1876 [astro-ph].

\bibitem{Yanagida:1987nf}
T.~Yanagida and M.~Yoshimura,
``Resonant axion-photon conversion in the Early Universe,''
Phys.\ Lett.\  B {\bf 202}, 301 (1988).

\bibitem{Fixsen:1996nj}
D.~J.~Fixsen, E.~S.~Cheng, J.~M.~Gales, J.~C.~Mather, R.~A.~Shafer and E.~L.~Wright,
``The Cosmic Microwave Background Spectrum from the Full COBE/FIRAS Data
Set,''
Astrophys.\ J.\  {\bf 473}, 576 (1996)
[astro-ph/9605054].

\bibitem{Mather:1998gm}
J.~C.~Mather, D.~J.~Fixsen, R.~A.~Shafer, C.~Mosier and D.~T.~Wilkinson,
``Calibrator Design for the COBE Far Infrared Absolute Spectrophotometer
(FIRAS),''
Astrophys.\ J.\  {\bf 512}, 511 (1999)
[astro-ph/9810373].

\bibitem{Jaeckel:2008fi}
J.~Jaeckel, J.~Redondo and A.~Ringwald,
``Signatures of a hidden cosmic microwave background,''
Phys.\ Rev.\ Lett.\  {\bf 101}, 131801 (2008)
[arXiv:0804.4157 [astro-ph]].

\bibitem{Mirizzi:2009iz}
A.~Mirizzi, J.~Redondo and G.~Sigl,
``Microwave Background Constraints on Mixing of Photons with Hidden
Photons,''
JCAP {\bf 0903}, 026 (2009)
[arXiv:0901.0014 [hep-ph]].

\bibitem{Mirizzi:2005ng}
A.~Mirizzi, G.~G.~Raffelt and P.~D.~Serpico,
``Photon axion conversion as a mechanism for supernova dimming: Limits from
CMB spectral distortion,''
Phys.\ Rev.\  D {\bf 72}, 023501 (2005)
[astro-ph/0506078].

\bibitem{Mirizzi:2007jd}
A.~Mirizzi, D.~Montanino and P.~D.~Serpico,
``Revisiting cosmological bounds on radiative neutrino lifetime,''
Phys.\ Rev.\  D {\bf 76}, 053007 (2007)
[arXiv:0705.4667 [hep-ph]].

\bibitem{Melchiorri:2007sq}
A.~Melchiorri, A.~Polosa and A.~Strumia,
``New bounds on millicharged particles from cosmology,''
Phys.\ Lett.\  B {\bf 650}, 416 (2007)
[hep-ph/0703144].

\bibitem{Raffelt:1996wa}
G.~Raffelt,
`` Stars as laboratories for fundamental physics,''
The University of Chicago Press (1996).

\bibitem{Masso:2005ym}
E.~Masso and J.~Redondo,
``Evading Astrophysical Constraints on Axion-Like Particles,''
JCAP {\bf 0509}, 015 (2005)
[hep-ph/0504202].

\bibitem{Jaeckel:2006xm}
J.~Jaeckel, E.~Masso, J.~Redondo, A.~Ringwald and F.~Takahashi,
``The Need for Purely Laboratory-Based Axion-Like Particle Searches,''
Phys.\ Rev.\  D {\bf 75}, 013004 (2007)
[hep-ph/0610203].

\bibitem{Masso:2006gc}
E.~Masso and J.~Redondo,
``Compatibility of CAST search with axion-like interpretation of PVLAS
results,''
Phys.\ Rev.\ Lett.\  {\bf 97}, 151802 (2006)
[hep-ph/0606163].

\bibitem{Brax:2007ak}
P.~Brax, C.~van de Bruck and A.~C.~Davis,
``Compatibility of the chameleon-field model with fifth-force experiments,
cosmology, and PVLAS and CAST results,''
Phys.\ Rev.\ Lett.\  {\bf 99}, 121103 (2007)
[hep-ph/0703243].

\bibitem{Mohapatra:2006pv}
R.~N.~Mohapatra and S.~Nasri,
Phys.\ Rev.\ Lett.\  {\bf 98}, 050402 (2007)
[arXiv:hep-ph/0610068].

\bibitem{Masso:1995tw}
 E.~Masso and R.~Toldra,
 ``On a Light Spinless Particle Coupled to Photons,''
 Phys.\ Rev.\  D {\bf 52}, 1755 (1995)
 [hep-ph/9503293].

\bibitem{Raffelt:1987im}
G.~Raffelt and L.~Stodolsky,
``Mixing of the Photon with Low Mass Particles,''
Phys.\ Rev.\  D {\bf 37}, 1237 (1988).

\bibitem{Wolfenstein:1977ue}
L.~Wolfenstein,
``Neutrino oscillations in matter,''
Phys.\ Rev.\  D {\bf 17}, 2369 (1978).
\bibitem{Mikheyev:1985aa}
S.~P. Mikheyev and A.~Y. Smirnov,
``Resonance enhancement of oscillations in matter and solar neutrino
spectroscopy,''
{\em Yad. Fiz.}
(1985) no.~42, 1441. [Sov.\
J.\ Nucl.\ Phys.\ {\bf 42}, 913 (1985)].
\bibitem{Mikheyev:1985bb}
S.~P. Mikheyev and A.~Y. Smirnov,
``Resonant amplification of neutrino oscillations in matter and solar
neutrino spectroscopy,''
{\em Nuovo Cimento C} (1986) no.~9, 17.
\bibitem{Parke:1986jy}
S.~J.~Parke,
``Nonadiabatic Level Crossing in Resonant Neutrino Oscillations,''
Phys.\ Rev.\ Lett.\  {\bf 57}, 1275 (1986).

\bibitem{Balantekin:1988aq}
 A.~B.~Balantekin, S.~H.~Fricke and P.~J.~Hatchell,
 ``Analytical And Semiclassical Aspects Of Matter Enhanced Neutrino
 Oscillations,''
 Phys.\ Rev.\  D {\bf 38}, 935 (1988).


\bibitem{Seager:1999bc}
S.~Seager, D.~D. Sasselov, and D.~Scott, ``{A New Calculation of the
Recombination Epoch},'' astro-ph/9909275.
\bibitem{Mack:2001gc}
A.~Mack, T.~Kahniashvili and A.~Kosowsky,
``Vector and Tensor Microwave Background Signatures of a Primordial
Stochastic Magnetic Field,''
Phys.\ Rev.\  D {\bf 65}, 123004 (2002)
[astro-ph/0105504].


\bibitem{Hu:1992dc}
W.~Hu and J.~Silk,
``Thermalization and spectral distortions of the cosmic background
radiation,''
Phys.\ Rev.\  D {\bf 48}, 485 (1993).

\bibitem{Iocco:2008va}
F.~Iocco, G.~Mangano, G.~Miele, O.~Pisanti and P.~D.~Serpico,
``Primordial Nucleosynthesis: from precision cosmology to fundamental
physics,''
Phys.\ Rept.\  {\bf 472}, 1 (2009)
[arXiv:0809.0631 [astro-ph]].

\bibitem{Masso:1997ru}
E.~Masso and R.~Toldra,
``New constraints on a light spinless particle coupled to photons,''
Phys.\ Rev.\  D {\bf 55}, 7967 (1997)
[hep-ph/9702275].

\bibitem{Kogut:1996zb}
  A.~Kogut,
  ``Diffuse Microwave Emission Survey,''
 astro-ph/9607100.

\bibitem{Fixsen:2002}
D.~J.~Fixsen, and J.~C.~Mather, 
``The Spectral Results of the Far-Infrared Absolute Spectrophotometer Instrument on COBE,''
ApJ {\bf 581} 817-822.

\bibitem{Singal:2008zz}
  J.~Singal,
  ``The CMB and galactic microwave absolute spectrum: Science and measurement
  with ARCADE 2,''
  Mod.\ Phys.\ Lett.\  A {\bf 23}, 1719 (2008).


\bibitem{Andriamonje:2007ew}
{\bf CAST} Collaboration, S.~Andriamonje {\it et al.},
``An improved limit on the axion-photon coupling from the CAST experiment,''
JCAP {\bf 0704}, 010 (2007)
[hep-ex/0702006].

\bibitem{Zioutas:2004hi}
{\bf CAST} Collaboration, K.~Zioutas {\it et al.},
``First results from the CERN Axion Solar Telescope (CAST),''
Phys.\ Rev.\ Lett.\  {\bf 94}, 121301 (2005)
[hep-ex/0411033].

\bibitem{Arik:2008mq}
{\bf CAST} Collaboration, E.~Arik {\it et al.},
``Probing eV-scale axions with CAST,''
JCAP {\bf 0902}, 008 (2009)
[arXiv:0810.4482 [hep-ex]].

\bibitem{Raffelt:2006cw}
G.~G.~Raffelt,
``Astrophysical axion bounds,''
Lect.\ Notes Phys.\  {\bf 741}, 51 (2008)
[hep-ph/0611350].

\bibitem{Brockway:1996yr}
J.~W.~Brockway, E.~D.~Carlson and G.~G.~Raffelt,
``SN 1987A gamma-ray limits on the conversion of pseudoscalars,''
Phys.\ Lett.\ B {\bf 383}, 439 (1996)
[astro-ph/ 9605197].

\bibitem{Grifols:1996id}
J.~A.~Grifols, E.~Mass\'o and R.~Toldr\`a,
``Gamma rays from SN~1987A due to pseudoscalar conversion,''
Phys.\ Rev.\ Lett.\ {\bf 77}, 2372 (1996)
[astro-ph/9606028].

\bibitem{Hoogeveen:1990vq}
F.~Hoogeveen and T.~Ziegenhagen,
``Production and detection of light bosons using optical resonators,''
Nucl.\ Phys.\  B {\bf 358}, 3 (1991).

\bibitem{Sikivie:2007qm}
P.~Sikivie, D.~B.~Tanner and K.~van Bibber,
``Resonantly enhanced axion - photon regeneration,''
Phys.\ Rev.\ Lett.\  {\bf 98}, 172002 (2007)
[hep-ph/0701198].

\bibitem{Cantatore:2008ju}
G.~Cantatore, R.~Cimino, M.~Karuza, V.~Lozza and G.~Raiteri,
``Polarization measurements and their perspectives: PVLAS Phase II,''
arXiv:0809.4208 [hep-ex].

\bibitem{ALPS}
 K.~Ehret {\it et al.},
 ``Resonant laser power build-up in ALPS -- a light-shining-through-walls experiment --'',
 arXiv:0905.4159 [physics.ins-det].

\bibitem{Fouche:2008jk}
{\bf BMV} Collaboration, M.~Fouche {\it et al.},
``Search for photon oscillations into massive particles,''
Phys.\ Rev.\  D {\bf 78}, 032013 (2008)
[arXiv:0808.2800 [hep-ex]].

\bibitem{Chou:2007zzc}
{\bf GammeV} Collaboration, A.~S.~Chou {\it et al.},
``Search for axion-like particles using a variable baseline photon regeneration technique,''
Phys.\ Rev.\ Lett.\  {\bf 100}, 080402 (2008)
[arXiv:0710.3783 [hep-ex]].

\bibitem{Afanasev:2008jt}
{\bf LIPSS} Collaboration, A.~Afanasev {\it et al.},
``New Experimental limit on Optical Photon Coupling to Neutral, Scalar Bosons,''
Phys.\ Rev.\ Lett.\  {\bf 101}, 120401 (2008)
[arXiv:0806.2631 [hep-ex]].

\bibitem{Battesti:2007zz}
R.~Battesti {\it et al.},
``The BMV experiment : a novel apparatus to study the propagation of light in a transverse magnetic field,''
Eur.\ Phys.\ Journ.\ D{\bf 46}, Issue 2,  pp.323-333 (2008),
arXiv:0710.1703v1 [physics.optics].

\bibitem{Khoury:2003rn}
J.~Khoury and A.~Weltman,
``Chameleon cosmology,''
Phys.\ Rev.\  D {\bf 69}, 044026 (2004)
[astro-ph/0309411].
\bibitem{Mota:2006ed}
D.~F.~Mota and D.~J.~Shaw,
``Strongly coupled chameleon fields: New horizons in scalar field theory,''
Phys.\ Rev.\ Lett.\  {\bf 97}, 151102 (2006)
[hep-ph/0606204].

\bibitem{Brax:2004qh}
P.~Brax, C.~van de Bruck, A.~C.~Davis, J.~Khoury and A.~Weltman,
``Detecting dark energy in orbit: The cosmological chameleon,''
Phys.\ Rev.\  D {\bf 70}, 123518 (2004)
[arXiv:astro-ph/0408415].
\bibitem{Mota:2003tm}
D.~F.~Mota and J.~D.~Barrow,
``Local and Global Variations of The Fine Structure Constant,''
Mon.\ Not.\ Roy.\ Astron.\ Soc.\  {\bf 349}, 291 (2004)
[arXiv:astro-ph/0309273].
\bibitem{Olive:2007aj}
K.~A.~Olive and M.~Pospelov,
``Environmental Dependence of Masses and Coupling Constants,''
Phys.\ Rev.\  D {\bf 77}, 043524 (2008)
[arXiv:0709.3825 [hep-ph]].



\bibitem{Zavattini:2007ee}
{\bf PVLAS} Collaboration, E.~Zavattini {\it et al.},
``New PVLAS results and limits on magnetically induced optical rotation and ellipticity in vacuum,''
Phys.\ Rev.\  D {\bf 77}, 032006 (2008)
[arXiv:0706.3419 [hep-ex]].

\bibitem{Ahlers:2007st}
M.~Ahlers, A.~Lindner, A.~Ringwald, L.~Schrempp and C.~Weniger,
``Alpenglow - A Signature for Chameleons in Axion-Like Particle Search Experiments,''
Phys.\ Rev.\  D {\bf 77}, 015018 (2008)
[arXiv:0710.1555 [hep-ph]].

\bibitem{Brax:2007hi}
P.~Brax, C.~van de Bruck, A.~C.~Davis, D.~F.~Mota and D.~J.~Shaw,
``Testing Chameleon Theories with Light Propagating through a Magnetic
Field,''
Phys.\ Rev.\  D {\bf 76}, 085010 (2007)
[arXiv:0707.2801 [hep-ph]].

\bibitem{Gies:2007su}
H.~Gies, D.~F.~Mota and D.~J.~Shaw,
``Hidden in the Light: Magnetically Induced Afterglow from Trapped Chameleon Fields,''
Phys.\ Rev.\  D {\bf 77}, 025016 (2008)
[arXiv:0710.1556 [hep-ph]].

\bibitem{Chou:2008gr}
{\bf GammeV} Collaboration, A.~S.~Chou {\it et al.},
``A Search for chameleon particles using a photon regeneration technique,''
Phys.\ Rev.\ Lett.\  {\bf 102}, 030402 (2009)
[arXiv:0806.2438 [hep-ex]].

\bibitem{Burrage:2008ii}
C.~Burrage, A.~C.~Davis and D.~J.~Shaw,
``Detecting Chameleons: The Astronomical Polarization Produced by
Chameleon-like Scalar Fields,''
Phys.\ Rev.\  D {\bf 79}, 044028 (2009)
[arXiv:0809.1763 [astro-ph]].
\bibitem{Burrage:2007ew}
C.~Burrage,
``Supernova Brightening from Chameleon-Photon Mixing,''
Phys.\ Rev.\  D {\bf 77}, 043009 (2008)
[arXiv:0711.2966 [astro-ph]].


\bibitem{Biggio:2006im}
C.~Biggio, E.~Masso and J.~Redondo,
``Mixing of photons with massive spin-two particles in a magnetic field,''
Phys.\ Rev.\  D {\bf 79}, 015012 (2009)
[hep-ph/0604062].

\bibitem{Deffayet:2000pr}
C.~Deffayet and J.~P.~Uzan,
``Photon mixing in universes with large extra-dimensions,''
Phys.\ Rev.\  D {\bf 62}, 063507 (2000)
[hep-ph/0002129].
\bibitem{De Angelis:2007yu}
A.~De Angelis, O.~Mansutti and M.~Roncadelli,
``Axion-Like Particles, Cosmic Magnetic Fields and Gamma-Ray Astrophysics,''
Phys.\ Lett.\  B {\bf 659}, 847 (2008)
[arXiv:0707.2695 [astro-ph]].
\bibitem{De Angelis:2007dy}
A.~De Angelis, O.~Mansutti and M.~Roncadelli,
``Evidence for a new light spin-zero boson from cosmological gamma-ray
propagation?,''
Phys.\ Rev.\  D {\bf 76}, 121301 (2007)
[arXiv:0707.4312 [astro-ph]].
\bibitem{DeAngelis:2008sk}
A.~De Angelis, O.~Mansutti, M.~Persic and M.~Roncadelli,
``Photon propagation and the VHE gamma-ray spectra of blazars: how
transparent is really the Universe?,''
arXiv:0807.4246 [astro-ph].
\bibitem{Csaki:2003ef}
C.~Csaki, N.~Kaloper, M.~Peloso and J.~Terning,
``Super-GZK photons from photon axion mixing,''
JCAP {\bf 0305}, 005 (2003)
[hep-ph/0302030].


\end{thebibliography}
\end{document}